\documentclass[a4papper,prd,onecolumn,superscriptaddress,nofootinbib,
eqsecnum,byrevtex,amsfonts,amssymb,amsmath]{revtex4}
\renewcommand{\vec}[1]{\boldsymbol{#1}}
\renewcommand{\exp}[1]{e^{#1}}
\newcommand{\ud}[1]{\mathrm{d}#1}
\newcommand{\br}[1]{\left(#1\right)}
\newcommand{\sq}[1]{\left[#1\right]}

\newcommand{\upd}[3]{{#1}^{#2}_{\phantom{#2}{#3}}}

\begin{document}
\title{van Stockum--Bonnor Spacetimes of Rigidly Rotating Dust}
\author{{\L}ukasz \surname{Bratek}}
\affiliation{Institute of
Nuclear Physics PAN, Radzikowskego 152, 31-342 Krak{\'o}w,
Poland.}
\author{Joanna \surname{Ja{\l}ocha}}
\affiliation{Institute of Nuclear Physics PAN, Radzikowskego
152, 31-342 Krak{\'o}w, Poland.}
\author{Marek \surname{Kutschera}}
\affiliation{Institute of Nuclear Physics PAN, Radzikowskego
152, 31-342 Krak{\'o}w, Poland.} \affiliation{Jagellonian
University, Institute of Physics, Reymonta 4, 30-059
Krak{\'o}w, Poland.}
\date{27 Feb 2007}
\begin{abstract}
Stationary, axisymmetric and asymptotically flat spacetimes
of dust of which trajectories are integral curves of the time
translation Killing vector are investigated. The flow has no
Newtonian limit. Asymptotic flatness implies the existence of
singularities of the curvature scalar that are distributions
and that are not isolated from regularity regions of the
flow. The singularities are closely related to the presence
of additional stresses that contribute negative active mass
to the total (Komar) mass, which is zero for asymptotically
flat spacetimes. Several families of solutions are
constructed.
\end{abstract}
\maketitle

\section{Introduction}
We study stationary, axisymmetric and asymptotically flat
spacetimes of dust in free fall along integral curves of the
time translation Killing vector. For reasons which will
become clear later, we shall call this motion 'van Stockum
flow'. Despite the flow is a rigid rotation with zero angular
velocity with respect to asymptotically static observers, the
squared vorticity scalar (which for the particular flow
equals, up to a constant factor, the proper energy
density of dust) in contrast to Newtonian
physics, does not vanish. Angular velocity of the flow with respect to locally
non-rotating observers numerically equals the angular
velocity of dragging of inertial frames, while angular velocity of matter in linearized
gravity is many orders of
magnitude greater. This shows that van Stockum flow is
ultra-relativistic even in the limit of negligible density.

We demonstrate that asymptotic flatness implies the existence
of curvature singularities that have distributional
character, spatial measure zero, and are not isolated
from regularity regions. The singularities are closely
related to additional weird stresses with negative active
mass. Total mass of such spacetimes is necessarily zero,
which has already been conjectured by Bonnor
\cite{bib:bonnorb}, and total angular momentum is determined
by the amplitude of the dipole in a multipole expansion.

We construct two classes of asymptotically flat solutions and
a class of asymptotically non-flat solutions. Bonnor's
solutions \cite{bib:bonnor},\cite{bib:bonnorb} belong to the
first two, and the van Stockum solution \cite{bib:vanstock} to
the third.

\section{van Stockum flow}
We shall focus on axisymmetric, stationary and asymptotically
flat spacetimes of dust flowing along opened integral
curves of the time translation Killing vector $\vec{\xi}$. By
asymptotic flatness the axial symmetry Killing vector
$\vec{\eta}$, of which integral lines are closed, vanishes on
the symmetry axis at least for radii sufficiently large.
Consequently, \textit{i)}
$\eta_{[\alpha}\xi_{\beta}\xi_{\mu;\nu]}=0$ and
$\xi_{[\alpha}\eta_{\beta}\eta_{\mu;\nu]}=0$ at least at one
point. The energy-momentum tensor is proportional to
$\xi_{\mu}\xi_{\nu}$, hence \textit{ii)}
$\xi^{\mu}R_{\mu}^{\phantom{\mu}[\nu}\xi^{\alpha}
\eta^{\beta]}=0$ and
$\eta^{\mu}R_{\mu}^{\phantom{\mu}[\nu}\xi^{\alpha}\eta^{\beta]}=0$
on the basis of Einstein's equations.  In addition, we assume
\textit{iii)} $[\vec{\xi},\vec{\eta}]=0$. Under these three
assumptions a theorem proved in \cite{bib:wald} guarantees
that, maybe apart from isolated points $(\rho,z)$, there
exist a coordinate system in which the line element takes the
general form
\begin{equation}
\ud{s}^2=-V\br{\ud{t}-K\ud{\phi}}^2+V^{-1}\rho^2\ud{\phi}^2
+\exp{2\Psi}\br{\ud{\rho}^2+\Lambda\ud{z}^2},
\label{eq:wald_metric}
\end{equation}
where $V$, $K$, $\Psi$ and $\Lambda$ are structure functions
of two variables $\rho$ and $z$. In these coordinates
$\vec{\xi}$ and $\vec{\eta}$ attain the particularly simple
form $\xi^{\mu}=\delta^{\mu}_t$ and
$\eta^{\mu}=\delta^{\mu}_{\phi}$.

The four-velocity of van Stockum flow reads
$\vec{u}=Z\vec{\xi}$, thus $Z^{-2}=-\xi^{\mu}\xi_{\mu}$ and
$\xi^{\mu}Z_{,\mu}=0$. Killing equations imply
$u_{(\mu;\nu)}=\xi_{(\mu}Z_{,\nu)}$, hence the expansion
scalar $u^{\mu}_{\phantom{\mu};\mu}=0$. On projecting
$u_{(\mu;\nu)}$ onto the $\vec{u}$-orthogonal subspace and
taking the traceless part, one infers the shear tensor
vanishes identically, as well. van Stockum flow is therefore
rigid. Despite the fact and that angular velocity of the
flow also vanishes ($\vec{u}\propto\vec{\xi}$), the vorticity
scalar $\omega^2$ does not, where
$$\omega^2=\frac{1}{4}\frac{\br{\vec{\xi}\vec{\xi}}^2
\br{\vec{\nabla}{S}}^2}{\br{\vec{\xi}\vec{\eta}}^2-\vec{\xi}^2\vec{\eta}^2},\qquad
S=\frac{\vec{\xi}\vec{\eta}}{\vec{\xi}\vec{\xi}}.$$

\subsection{Equations of van Stockum flow}
By definition of the flow, the energy-momentum tensor reads
$T_{\mu\nu}=\mathcal{D}Z^2\xi_{\mu}\xi_{\nu}$, where
$\mathcal{D}$ is the proper energy density such that
$\xi^{\mu}\mathcal{D}_{,\mu}=0=\eta^{\mu}\mathcal{D}_{,\mu}$.
Einstein's equations and the contracted Bianchi identity
imply the flow is continuous and geodesic. As
$u^{\mu}_{\phantom{\mu};\mu}=0$, the continuity equation
$\br{\mathcal{D}Z\xi^{\mu}}_{;\mu}=0$ is satisfied
identically. In addition,
$u^{\nu}u_{\mu;\nu}=-u^{\nu}Z\xi_{\nu;\mu}$. The geodesic
equation $u^{\nu}u_{\mu;\nu}=0$ will be satisfied if
$0=-u^{\nu}u_{\nu;\mu}+u^{\nu}u_{\nu}\br{\ln{Z}}_{,\mu}$. As
$u^{\mu}u_{\mu}\equiv-1$, $Z$ must be constant. This, in
turn, implies $V$ is also constant and, without loss of
generality, we may set $V\equiv1$. On defining
$K^{\mu\nu}=\vec{\xi}^2\eta^{\mu}\eta^{\nu}+
2\vec{\xi}\vec{\eta}\xi^{\mu}\eta^{\nu}+
\vec{\eta}^2\xi^{\mu}\xi^{\nu}$ we obtain for dust
$K^{\mu\nu}\br{T_{\mu\nu}-Tg_{\mu\nu}/2}=0$, then Einstein's
equations imply $K^{\mu\nu}R_{\mu\nu}=0$ or
$\rho\exp{-2\Psi}\partial_{\rho}\ln{\sqrt{|\Lambda|}}=0$ in
coordinates, hence $\Lambda=\Lambda(z)$. If so, the form of
(\ref{eq:wald_metric}) allows us to set $\Lambda(z)\equiv1$.
We have thus shown that the line element of van Stockum flow
reads
\begin{equation}
\ud{s}^2=-\ud{t}^2+2K(\rho,z)\ud{t}\ud{\phi}+(\rho^2-K^2(\rho,z))\ud{\phi}^2
+\exp{2\Psi(\rho,z)}
\br{\ud{\rho}^2+\ud{z}^2}\label{eq:vs_metric}.\end{equation}
Let $\upd{E}{\mu}{\nu}=\upd{R}{\mu}{\nu}-
\frac{1}{2}R\upd{\delta}{\mu}{\nu}-8\pi\upd{T}{\mu}{\nu}$,
then $\upd{E}{\rho}{\rho}=0$ and $\upd{E}{\rho}{z}=0$ yield the following relations
\begin{equation}\Psi_{,\rho}=\frac{K_{,z}^2-K_{,\rho}^2}{4\rho}, \qquad
\Psi_{,z} =-\frac{K_{,\rho}K_{,z}}{2\rho}.
\label{eq:nu}\end{equation} The integrability condition
$\Psi_{,\rho{z}}=\Psi_{,z\rho}$ imposes on $K$ the elliptic
constraint
\begin{equation}\mathcal{L}K=0, \qquad \mathcal{L}=
\partial^2_{\rho}-\frac{1}{\rho}\partial_{\rho}+\partial^2_{z}.
\label{eq:n}\end{equation} Provided (\ref{eq:nu}) and
(\ref{eq:n}) are satisfied, the other components of
$\upd{E}{\mu}{\nu}$, but $\upd{E}{t}{t}$ and
$\upd{E}{t}{\phi}$, vanish identically. The latter two will
also vanish for $\mathcal{C}^2$ solutions (we stress, the reservation 'for $\mathcal{C}^2$ solutions' is necessary)
if only
$$\mathcal{D}=\exp{-2\Psi}
\frac{K_{,\rho}^2+K_{,z}^2}{8\pi\rho^2}.
$$

\section{Asymptotic flatness and curvature singularities}

A function $K^{\epsilon}$ is a regularized profile of a solution
$K$ of (\ref{eq:n}) in an open subset $\mathcal{V}$ of the
plane $(\rho,z)$ if
$K^{\epsilon}\in\mathcal{C}^{\infty}(\mathcal{V})$ and
$K^{\epsilon}\to{}K$ as $\epsilon\to0$ almost everywhere in $\mathcal{V}$. We
shall denote by $\mathcal{I}$ the set of all irregularity
points of van Stockum flow, that is, a subset of
$\mathbb{R}^3$ where
$\mathcal{L}K$ does not exist for solutions in the sense that
$\lim_{\epsilon\to0}\int_{\mathcal{\mathcal{I}_{\delta}}}
\ud{m}f\mathcal{L}K^{\epsilon}\ne0$ for any $\delta>0$, where
$\mathcal{I}\subset\mathcal{I}_{\delta}$,
$0<dist(\partial\mathcal{I}_{\delta},\mathcal{I})<\delta$,
$\exp{2\Psi}f=\rho^{-1}K^{\epsilon}_{,\rho}$ or
$\rho^{-2}K^{\epsilon}$ and 
$\ud{m}=\exp{2\Psi}\rho\ud{\rho}\ud{\phi}\ud{z}$. In particular,
$K\not\in\mathcal{C}^2(\mathcal{I})$. We recall the result of
the theory of elliptic equations that
$\mathcal{I}/\mathcal{S}^1$ has measure zero in the plane
$(\rho,z)$.

In what follows we shall prove that $\mathcal{I}\ne\emptyset$
for asymptotically flat van-Stockum spacetimes. For if we
suppose for contradiction that $\mathcal{I}=\emptyset$, then
inside a ball $\mathcal{B}_{\textsf{R}}\subset\mathbb{R}^3$
bounded by a two-sphere $\mathcal{S}_{\textsf{R}}$ of radius
$R$ and concentric with the origin
\begin{equation}\label{eq:stokes}\int\limits_{\mathcal{B}_{\textsf{R}}}
\mathcal{D}\exp{2\Psi}\rho\ud{\rho}
\wedge\ud{\phi}\wedge\ud{z}\stackrel{\stackrel{\mathcal{L}K=0}{}}{=}
\frac{1}{8\pi}\int\limits_{\mathcal{S}_{\textsf{R}}}
\frac{K}{\rho}\br{K_{,z}\ud{\rho}-K_{,\rho}\ud{z}
}\wedge\ud{\phi}\equiv\frac{1}{8\pi}\int\limits_{
\mathcal{S}_{\textsf{R}}}\frac{K\partial_rK}{\sin{\theta}}
\ud{\phi}\wedge\ud{\theta}\end{equation} in virtue of the
Stokes theorem, provided $(K^2)_{,r}=o(\sin{\theta})$,
($r\sin{\theta}=\rho$, $r\cos{\theta}=z$). By asymptotic
flatness $K\sim2Jr^{-1}\sin^2{\theta}$ as $r\to\infty$,
hence, for $\textsf{R}$ sufficiently large, the surface
integral on the right-hand side is negative and tends to $0$
as $\textsf{R}\to\infty$, while the volume integral is
positive, a contradiction, thus indeed
$\mathcal{I}\ne\emptyset$.

The surface integral (\ref{eq:stokes}) coincides in the limit
$\mathsf{R}\to\infty$ with the total (Komar) mass $M$, which
is zero. Indeed, for asymptotically flat spacetimes with 
metric ($\ref{eq:vs_metric}$), $M$ is given by\footnote{The
$2$-forms in (\ref{eq:gauss_m}) are equal on
$\mathcal{S}_{\textsf{R}}$ irrespective of whether Einstein's
equations are satisfied or not}
\begin{equation}\label{eq:gauss_m}M=-\frac{1}{8\pi}\lim_{\textsf{R}\to\infty}
\int\limits_{\mathcal{S}_{\textsf{R}}}\frac{1}{2}\sqrt{-g}
\epsilon_{\alpha\beta\mu\nu}\nabla^{\mu}\xi^{\nu}
\ud{x}^{\alpha}\wedge\ud{x}^{\beta}=\lim_{\textsf{R}\to\infty}\frac{1}{8\pi}\int\limits_{
\mathcal{S}_{\textsf{R}}}\frac{K\partial_rK}{\sin{\theta}}
\ud{\phi}\wedge\ud{\theta}=0.\end{equation} We conclude,
therefore, that the integral $\lim_{\textsf{R}\to\infty}
\int_{\mathcal{B}_{\textsf{R}}}\ud{m}(8\pi\rho^2)^{-1}
\exp{-2\Psi}K\mathcal{L}K$, which was omitted deliberately in
(\ref{eq:stokes}) as it would be zero for $\mathcal{C}^2$
solutions, does not vanish. For regularized
profiles $K^{\epsilon}$ this integral
tends in the limit $\epsilon\to0$ to minus the total mass
$\int_{\mathbb{R}^3\setminus\mathcal{I}}\mathcal{D}\ud{m}$ of
the regularity region $\mathbb{R}^3\setminus\mathcal{I}$.
Putting this in other words, asymptotically flat van Stockum
spacetimes contain additional sources of negative active mass
located in $\mathcal{I}$. The sources are distributions as
they have both finite mass and a spatial support of measure
zero.

To be more explicit, the proper energy density
$\widetilde{\mathcal{D}}=T_{\mu\nu}u^{\mu}u^{\nu}$, the trace
of spatial stresses
$\widetilde{\mathcal{S}}=T_{\mu\nu}\br{u^{\mu}u^{\nu}+g^{\mu\nu}}$,
Tolman's active mass density on a hypersurface of constant
time
$\widetilde{D}_{T}=\br{8\pi}^{-1}R^{t}_{\phantom{t}\mu}\xi^{\mu}$,
and the curvature scalar
$\widetilde{R}=R^{\mu}_{\phantom{\mu}\mu}$ of a spacetime
with metric (\ref{eq:vs_metric}) read
\begin{eqnarray*}&\widetilde{\mathcal{D}}=
\frac{3}{4}\exp{-2\Psi}
\frac{K_{,\rho}^2+K_{,z}^2}{8\pi\rho^2}-
\frac{1}{8\pi}\exp{-2\Psi}\br{\Psi_{,\rho\rho}+
\Psi_{,zz}},\quad \widetilde{\mathcal{D}}_{T}=\exp{-2\Psi}
\frac{K_{,\rho}^2+K_{,z}^2}{8\pi\rho^2}+
\frac{1}{8\pi}\exp{-2\Psi}\frac{K}{\rho^2}\mathcal{L}K,&
\\
&\widetilde{\mathcal{S}}=\frac{1}{4}\exp{-2\Psi}
\frac{K_{,\rho}^2+K_{,z}^2}{8\pi\rho^2}+\frac{1}{8\pi}\exp{-2\Psi}
\br{\Psi_{,\rho\rho}+ \Psi_{,zz}},\qquad
\widetilde{\mathcal{R}}
=8\pi(\widetilde{\mathcal{D}}-\widetilde{\mathcal{S}}).&
\end{eqnarray*} For any smooth $\Psi$ and $K$
satisfying only (\ref{eq:nu}), the above definitions reduce
to
\begin{eqnarray*}
&\widetilde{\mathcal{D}}=\mathcal{D}+\exp{-2\Psi}
\frac{K_{,\rho}}{16\pi\rho}\mathcal{L}K,\quad
\widetilde{\mathcal{D}}_{T}=\mathcal{D}+
\exp{-2\Psi}\frac{K}{8\pi\rho^2}\mathcal{L}K,&\\
&
\widetilde{\mathcal{S}}=-\exp{-2\Psi}\frac{K_{,\rho}}{16\pi\rho}
\mathcal{L}K,\quad
\widetilde{R}=8\pi\mathcal{D}+\exp{-2\Psi}\frac{K_{,\rho}}{\rho}
\mathcal{L}K.&
\end{eqnarray*}
In particular, for regularized profiles $K^{\epsilon}$
 we obtain, outside $\mathcal{I}_{\delta}$ and in the limit
$\epsilon\to0$,
$\widetilde{\mathcal{D}}=\widetilde{\mathcal{D}}_T=D=
\br{8\pi}^{-1}\widetilde{R}$ and $\widetilde{\mathcal{S}}=0$,
like for dust. However, in the same limit
$\int_{\mathcal{I}_{\delta}}\widetilde{S}\ud{m}\ne0$ for
arbitrarily small $\delta$. Note, that $\mathcal{I}$ is the
scalar curvature singularity as $\widetilde{R}$
is a distribution on $\mathcal{I}$. Indeed, if $K_{\epsilon}$
is a regularized profile of an asymptotically flat solution
then $\int_{\mathbb{R}^3}\widetilde{R}\ne
\int_{\mathbb{R}^3\setminus\mathcal{I}}\widetilde{R}=
8\pi\int_{\mathbb{R}^3}\mathcal{D}$ in the limit
$\epsilon\to0$. Thus
$\widetilde{R}=8\pi\mathcal{D}+\gamma_{\mathcal{I}}$ where
$D$ is smooth and integrable, and $\gamma_{\mathcal{I}}$ is a
distribution localized on $\mathcal{I}$. The singularity is
not isolated from regularity regions.\footnote{In this sense
we cannot agree with the statement of paper
\cite{bib:bonnorb} that a solution found therein had no
curvature singularity.}

Here we give an example illustrating the statements. The
Bonnor solution \cite{bib:bonnorb} with an embedded surface
layer of negative mass, and which we shall denote by
$K^0_{B}$, can be regularized by defining
$K^{\epsilon}_B(\rho,z)=
\sqrt{8a^3\mu}\cdot\rho^2\cdot\br{(a+\sqrt{z^2+\epsilon^2})^2+
\rho^2}^{-3/2}$, $a>0$, $\mu>0$. Although $K^{\epsilon}_B$ is
globally $\mathcal{C}^\infty$, its limit $K^{0}_B$ is not
even differentiable in $\mathcal{I}$, which is the plane
$z=0$. On integrating over $\mathbb{R}^3$ and taking the
limit $\epsilon\to0$ we obtain
$\int_{\mathbb{R}^3}D\ud{m}=\int_{\mathbb{R}^3\setminus\mathcal{I}}D\ud{m}=\mu$,
$\int_{\mathbb{R}^3}\widetilde{D}\ud{m}=3\mu/4$,
$\int_{\mathcal{I}}\widetilde{S}\ud{m}=\mu/4$,
$\int_{\mathbb{R}^3}\widetilde{R}\ud{m}=4\pi\mu$
 and
$\int_{\mathbb{R}^3}\widetilde{D}_{T}\ud{m}\equiv0$. The
latter holds identically as
$8\pi\sqrt{-g}\widetilde{D}_{T}\ud{\rho}\wedge\ud{\phi}\wedge\ud{z}=
\ud\br{\rho^{-1}KK_{,z}\ud{\rho}\wedge\ud{\phi}+\rho^{-1}
KK_{,\rho}\ud{\phi}\wedge\ud{z}}$ for $\mathcal{C}^2$
functions, hence
$\int_{\mathbb{R}^3}\widetilde{D}_{T}\ud{m}\equiv{}M=0$ for
regularized asymptotically flat profiles. Since for $K^0_B$ $8\pi{D}=\widetilde{R}$ only outside $\mathcal{I}$,
and since $\int_{\mathbb{R}^3\setminus\mathcal{I}}D\ud{m}=
\int_{\mathbb{R}^3}D\ud{m}=\mu\ne\mu/2=
\br{8\pi}^{-1}\int_{\mathbb{R}^3}\widetilde{R}\ud{m}$, the
curvature scalar is a distribution, and $\widetilde{R}$ is smooth
and bounded only outside $\mathcal{I}$.

These properties and those discussed in the introduction show
that global and asymptotically flat van Stockum spacetimes
are not viable physically. The same, for example, concerns
also asymptotically flat stationary and axisymmetric
spacetimes of which internal metrics are matched onto the
external line element (\ref{eq:vs_metric}), as then $M=0$.
However, a possibility that van Stockum spacetime can be part
of a regular spacetime cannot be excluded.

\section{Three classes of solutions}
\subsection{Solutions with a layer of negative mass}
Solutions to equation (\ref{eq:n}) can be sought via integral
transforms, for example,
$$K(\rho,z)=\rho\int\limits_{0}^{\infty}
\lambda\hat{K}_{\textsf{J}}(\lambda)\exp{-\lambda
|z|}\textsf{J}_{1}(\lambda\rho)\ud{\lambda}, \quad
\mathrm{or} \quad \rho\int\limits_{0}^{\infty}\lambda
\hat{K}_{\textsf{K}}(\lambda)
\cos{(\lambda{z})}\textsf{K}_{1}(\lambda\rho)\ud{\lambda}
$$
generate $z$--symmetric solutions; $\textsf{J}_1$ and
$\textsf{K}_1$ are Bessel functions. To give an example, the
solution  $\rho^2r^{-3}$ discussed in \cite{bib:bonnor} has
$\hat{K}_{\textsf{J}}(\lambda)=1$ and
$\hat{K}_{\textsf{K}}(\lambda)=2/\pi$, while the Bonnor 
solution $K^0_B$ has
$\hat{K}_{\textsf{J}}(\lambda)\propto{}\exp{-a\lambda}$. The
latter belongs to a class of solutions defined by
specifying
$\hat{K}_{\textsf{J}}(\lambda)=\frac{l^{2n+2}\lambda^{2n}}{(2n+1)!}\exp{-a\lambda}$,
$n\in\mathbb{N}$, which yields
$$K_n(\rho,z)=\frac{(n+1)l^{2n+2}\rho^2}{(a+|z|)^{2n+3}}
\cdot{}_2F_1\br{\frac{3}{2}+n,2+n;2,-\frac{\rho^2}{(a+|z|)^2}}.$$
Apart from the plane $z=0$ the solutions are smooth
everywhere. One can show that
$|K_n|<\rho\br{l/\br{a+|z|}}^{2n+2}$, thus, at least for $l<a$,
hypersurfaces of constant $t$ are globally space-like as
then $|K_n|<\rho$. Since
$K_n\partial_rK_n(r,\theta)\sim{}r^{-(4n+3)}\sin^4{\theta}$
times a bounded geometrical factor, the spacetimes are
asymptotically flat with $M=0$. Function
$\mathcal{D}\exp{2\Psi}$ is finite for $z\ne0$ and for $r$
sufficiently large behaves as $r^{-(4n+6)}$. The plane $z=0$
is thus a curvature singularity with finite and negative
active mass. Only the Bonnor solution $K^{0}_B$ $(n=0)$ has
nonzero angular momentum.

\subsection{External and Internal multipolar solutions and a multipole expansion} Let
$K(\rho,z)=W(r)Y(\cos{\theta})$, where $\rho=r\sin{\theta}$,
$z=r\cos{\theta}$. There exist three families of solutions to
equation (\ref{eq:n}) satisfying $r^2W''(r)=\lambda{}W(r)$
and $(1-x^2)Y''(x)+\lambda{}Y(x)=0$, ($x=\cos{\theta}$); with
\textit{i)} $\lambda=\alpha(\alpha+1)$, $\alpha\geq0$;
\textit{ii)} $-\cos^2(\alpha)/4$, $0\leq\alpha<\pi/2$ and
\textit{iii)} $-\cosh^2(\alpha)/4$, $\alpha>0$. The
\textit{i)} class contains $x$-analytic external ($W=r^{-n}$)
and internal ($W=r^{n+1}$) solutions. In this way we obtain
external $K^{(n)}_E$ and internal $K^{(n)}_I$ multipolar
solutions
\begin{table}[h!]
\begin{tabular}{|c|c|c|}
\hline
$n$&$0,2,4,6,...$&$1,3,5,7,...$\\
\hline $K_E^{(n)}(\rho,z)$&
$\frac{z}{\br{\rho^2+z^2}^{\frac{n+1}{2}}}A_n
\br{\frac{z^2}{\rho^2+z^2}}$&$\frac{1}{\br{\rho^2+z^2}^{n/2}}
B_n\br{\frac{z^2}{\rho^2+z^2}}$\\
\hline $K_I^{(n)}(\rho,z)$ & $z\br{\rho^2+z^2}^{n/2}
A_n\br{\frac{z^2}{\rho^2+z^2}}$& $\br{\rho^2+z^2}^{(n+1)/2}
B_n\br{\frac{z^2}{\rho^2+z^2}}$\\
\hline
\end{tabular}
\end{table}
\\ where $A_n(y)={}
_2F_1\br{\frac{1}{2}+\frac{n}{2},-\frac{n}{2};\frac{3}{2},y}$
and $B_n(y)={}
_2F_1\br{-\frac{1}{2}-\frac{n}{2},\frac{n}{2};\frac{1}{2},
\frac{z^2}{\rho^2+z^2}}$. Internal solutions $K_I^{(n)}$ (of
which element is the van Stockum solution
\cite{bib:vanstock}) give rise to spacetimes that are not
asymptotically flat. With the exception of the monopole
$K_E^{(0)}$, $K_E^{(n)}$ yield asymptotically flat
spacetimes that contain in the
center pathological singularities with non-integrable $\mathcal{D}$ (e.g. $\exp{2\Psi}\mathcal{D}>a^4/\br{2\pi{}r^6}$
for the dipole $K_E^{(1)}=a^2\sin^2{(\theta)}r^{-1}$). For the solutions $M=0$, thus contributions to
$M$ from the singularities are formally $-\infty$. Another
non-physical property of $K_E^{(n)}$ is that
$|K(\rho,z)|>\rho$ in the vicinity of the center. In the
region the axial symmetry Killing vector $\vec{\eta}$ is
time-like, as so, the region contains closed time-like
curves (e.g. $K_E^{(1)}$ is such inside the
region bounded by $\rho=a\sqrt{\sin{\alpha}}\sin{\alpha}$,
$z=a\cos{\alpha}\sqrt{\sin{\alpha}}$, $\alpha\in(0,\pi)$).
However, external multipoles $K_E^{(n>0)}$ appear in
multipolar expansions of asymptotically flat solutions. We
illustrate this by giving an example below.

It should be clear that
$K_a=-\int_{-a}^{a}a^{-1}{s}\ud{s}f(\rho,z-s)$, $a>0$, where
$f(\rho,z)=K^{(0)}_E=zr^{-1}$, is a $z$-symmetric solution such that $0\leq{}K_a\leq{}a$. The conformal mapping $z+i\rho=a\cosh{\br{u+iv}}$
($u\geq0$, $0\leq{}v\leq\pi$) is invertible apart from two points
$(\rho,z)=(0,\pm{a})$ and
transforms $S_a=\{(\rho,z):\rho=0, z\in[-a,a]\}$ to the
segment $u=0$, $v\in[0,\pi]$. In this map the solution reads
$$
K_a(u,v)=a\sin^2(v)\br{\cosh(u)+\frac{1}{2}\sinh^2(u)
\ln\sq{\tanh^2\br{\frac{u}{2}}}} \label{eq:my_sol}.$$ The
resulting spacetime is asymptotically flat as
$K_a\sim(4/3)a\exp{-u}\sin^2{v}$ for large $u$. In the
vicinity of $u=0$, $K_a\sim{}u^2\ln{u}$, therefore
$\mathcal{I}=\mathcal{S}_a$. Multipolar expansion of $K_a$ in
the base of functions $K^{(n)}_E$, $n=1,3,5,\dots$, reads
$$K_{a}(\rho,z)=\frac{2}{3}\frac{\rho^2}{r^3}a^2-\frac{1}{5}
\frac{\rho^2\br{\rho^2-4z^2}}{r^7}a^4+\frac{3}{28}\frac{\rho^2
\br{\rho^4-12\rho^2z^2+8z^4}}{r^{11}}a^6+\dots,$$ and
$\Psi_a(r,\theta)\sim-(a^4/72)r^{-4}\br{7+9\cos{2\theta}}
\sin^2{\theta}$ as $r\to\infty$. Asymptotically, (\ref{eq:vs_metric})
reduces to
$$\ud{s}^2\sim-\ud{t}^2-\frac{4}{3}\frac{a^2}{r}\sin^2{\theta}
\ud{t}\ud{\phi}+\ud{r}^2
+r^2\br{\ud{\theta}^2+\sin^2{\theta}\ud{\phi^2}}.$$
Comparison with the asymptotic expansion of the Kerr metric
gives total mass $M=0$ and total angular momentum $J=a^2/3$
in agreement with (\ref{eq:gauss_m}) and with the analogous
expression for the total angular momentum of asymptotically flat
van Stockum flow $$J=\lim_{r\to\infty}\frac{1}{16\pi}
\iint\sq{\frac{2K}{r}- \br{1+\frac{K^2}{r^2\sin^2{\theta}}}
\partial_rK}r^2\sin{\theta}\ud{\theta}\ud{\phi}.
$$


\end{document}